\documentclass[aps,pra,showpacs,notitlepage]{revtex4-1}

\usepackage{graphicx}

\begin{document}

\title{Control of particle propagation beyond the uncertainty limit\\
by interference between position and momentum}

\author{Holger F. Hofmann}
\email{hofmann@hiroshima-u.ac.jp}
\affiliation{
Graduate School of Advanced Sciences of Matter, Hiroshima University,
Kagamiyama 1-3-1, Higashi Hiroshima 739-8530, Japan
}

\begin{abstract}
As shown in Phys. Rev. A {\bf 96}, 020101(R) (2017), it is possible to demonstrate that quantum particles do not move along straight lines in free space by increasing the probability of finding the particles within narrow intervals of position and momentum beyond the ``either/or'' limit of 0.5 using constructive quantum interference between a component localized in position and a component localized in momentum. The probability of finding the particle in a corresponding spatial interval at a later time then violates the lower bound of the particle propagation inequality which is based on the validity of Newton's first law. In this paper, the problem of localizing the two state components in their respective target intervals is addressed by introducing a set of three coefficients that describe the localization of arbitrary wavefunctions quantitatively. This characterization is applied to a superposition of Gaussians, obtaining a violation of the particle propagation inequality by more than 5 percent if the width of the Gaussian wavefunction is optimized along with the size of the position and momentum intervals. It is shown that the violation of the particle propagation inequality originates from the fundamental way in which quantum interferences relate initial position and momentum to the future positions of a particle, indicating that the violation is a fundamental feature of causality in the quantum limit. 
\end{abstract}

\maketitle

\section{Introduction}

Quantum mechanics replaces the precise concepts of particle position and particle momentum with the rather less clear concept of a quantum state that describes only the statistical distributions of possible particle positions and particle momenta. Since statistical statements are by their very nature incomplete, there is much room left for speculation regarding the quantum limit of motion for individual point-like particles \cite{War02,Pfi04,Hal13a,Saw14,Mag16,Zho17}. Interestingly, very little of the discussion has focused on the simplification of quantitative criteria for the evaluation of particle propagation, and most of the proposed experimental approaches involve quite a bit of complicated data analysis \cite{Koc11,Lun11,Sch13,Bli13,Hal14,Mor16}. Yet it seems obvious that quantum interference effects significantly modify the motion of particles even in free space. For instance, it has been observed that the probability currents in the time evolution of particles in free space can flow backwards even when the probability of negative momentum is virtually zero. Unfortunately, this quantum backflow effect is rather difficult to isolate and clear experimental evidence may be hard to obtain \cite{Bra94,Hal13b}. It might well be that the problem with much of the previous discussion is an element of circular reasoning by which we automatically associate motion with differential shifts in time. However, such differential shifts in time are difficult to define in quantum mechanics, since no realistic state or measurement achieves the necessary precision, and the theoretical assumption of zero position uncertainty at one time always implies infinite uncertainties at all other times. It may therefore be necessary to change our perspective on the problem: instead of formulating the laws of motion in terms of differential shifts and their associated probability currents, we should consider the large scale patterns that relate the statistics of positions observed at different times to each other, where the time difference $t$ should be large enough to unambiguously identify changes in position even in the presence of unavoidable position and momentum uncertainties. 

As has been shown recently, it is possible to derive a statistical limit for free space propagation in a straight line that refers only to initial position, initial momentum, and a second position obtained at a later time $t$ \cite{Hof17}. The advantage of this approach is that it relates to the practical problem of controlling the position at time $t$ by choosing the appropriate combination of position and momentum at time zero. Intuitively, it would seem obvious that a simultaneous control of position and momentum would allow us to control the complete path of a particle, so that any trick by which we could improve the joint definition of initial position and momentum would have to result in a corresponding improvement of the probability of finding the particle at the target position at time $t$. However, quantum mechanics proves this expectation wrong. As pointed out in \cite{Hof17}, quantum interference between a state with a rectangular wavefunction confined in a position interval of width $L$ and a corresponding state confined in a momentum interval of width $B$ increases the probabilities of finding the particle in these intervals beyond fifty percent, resulting in a minimal joint probability of position and momentum that allows us to test whether this joint control of position and momentum results in the arrival of the particle at the intended position at a later time $t$. Specifically, we can compare the minimal joint probability obtained from the separate measurements of position and momentum with the actual probability of finding the particle in the corresponding position interval at time $t$. In the case of quantum interferences between position and momentum, the probability observed at time $t$ can be lower than the minimal joint probability, and this violation of Newton's first law can be quantified by the defect probability which is defined as the difference between the observed probability and the minimal joint probability determined from the separate distributions of position and momentum. The problem can be optimized by choosing the propagation time $t=mL/B$, where the contributions of the momentum uncertainty $B$ to the position uncertainty is exactly equal to the contribution of the initial position uncertainty $L$. It is then possible to show that the probability of finding the particle in the target interval can be significantly lower than the minimal joint probability of initial position and momentum, with a maximal defect probability of more than seven percent for an optimized interval product of $LB=0.024 (2 \pi \hbar)$ \cite{Hof17}. 

In the initial analysis, rectangular wavefunctions were used to achieve perfect confinement within the intended position and momentum intervals. Projections on these rectangular states correspond to the detection of a particle within the respective interval, providing an easy method of evaluating the relevant probabilities involved in the particle propagation paradox. Rectangular wavefunctions also appear to be an optimal choice for the simultaneous control of position and momentum, since their interference terms are fully localized within both the position and the momentum intervals. Moreover, the preparation of a rectangular wavefunction is not uncommon in optical systems, as demonstrated by experiments on spatial qubits \cite{Nev07,Tag08,Pee09,Sol11}. However, it may be important to test whether the observed effect is robust against changes of the wavefunction, and how wavefunctions that are not perfectly confined in the respective intervals change the statistics used to demonstrate the failure of Newton's first law in \cite{Hof17}. In the following, I will therefore consider the violation of the statistical limits for free space propagation along straight lines for a wider class of superposition states, evaluating the localization in the position and momentum intervals for arbitrary wavefunctions. In particular, it is shown that superpositions of two Gaussian wavefunctions can achieve a defect probability of more than five percent, a result that can be predicted using an optimization based on only a few characteristic coefficient describing the localization of the wavefunctions in the superposition. 

The rest of the paper is organized as follows. In Sec. \ref{Sec:uncertainty}, the problem of controlling particle propagation in the presence of statistical uncertainties is introduced. In Sec. \ref{Sec:Gauss}, it is shown that a defect probability of more than five percent can be obtained using a superposition of two Gaussian wavefunctions. The particle propagation paradox is therefore robust against the specific waveform and should be observable with a wide variety of superpositions. In Sec. \ref{Sec:coeff}, the localization of a wavefunction in the intervals L and B is characterized by introducing appropriate coefficients. In Sec. \ref{Sec:control}, the superposition of a state localized in position and a state localized in momentum is introduced and its localization statistics are discussed. In Sec. \ref{Sec:prop}, the probability distribution is analyzed at a propagation time of $t=mL/B$ where the two components of the state have equal width and approximately equal shapes. It is shown that quantum interferences between position and momentum generally result in a lower probability than the one required by propagation in a straight line. An estimate for the minimal defect probability describing the violation of the particle propagation inequality is derived for the coefficients that characterize the localization of the quantum state components. In Sec. \ref{Sec:LocalG}, the characteristic coefficients of Gaussian wavefunctions are derived. The optimized result corresponds to the more precise calculation in Sec. \ref{Sec:Gauss}. In Sec. \ref{Sec:causality}, the relation between quantum interference and causality discussed and the possibility that quantum interference may provide the correct microscopic explanation of all causality relations is considered \cite{Hof12,Hof14}. Finally, Sec. \ref{Sec:concl} summarizes the results and concludes the paper.

\section{Uncertainties in the control of particle propagation}
\label{Sec:uncertainty}

As mentioned in the introduction, the scenario considered in the following concerns the possibility of controlling the trajectory of a particle in free space by jointly controlling the initial position and the initial momentum. The mathematical formalism of quantum theory seems to suggest that such a joint control of position and momentum would result in the corresponding propagation along a straight line since the Heisenberg equations of motion describe the time dependence of the position operator in the same manner that is suggested by Newton's first law,   
\begin{equation}
\label{eq:Heis}
\hat{x}(t) = \hat{x}(0) + \frac{1}{m} \hat{p}(0) t.
\end{equation}
The obvious problem with this equation is that it refers to operators, and not to individual values of position and momentum. This means that we are unable to assign specific values to the physical properties, except in the special case of eigenstates. The Hilbert space formalism itself thus dictates an uncertainty trade-off between the initial position and the initial momentum that will be carried over into the quantum statistics of future positions $\hat{x}(t)$. 

Any practical scenario of control must deal with the problem of statistical uncertainty in a suitable quantitative manner. The traditional approach to uncertainty uses the variance of the distribution as a quantitative measure, but this is not a useful measure if the intention is to derive precise constraints on the properties of individual systems. For this purpose, it is necessary to consider the actual probabilities of obtaining specific measurement results instead. This alternative concept of a statistical uncertainty is close to the approach that has been developed within the frameworks of joint measurements and of entropic uncertainty relations \cite{Lah86,Bus07,Rud15,Col17}. In the following, joint control over position and momentum will be defined in terms of the probability of finding the initial position in an interval of width $L$ and the probability of finding the initial momentum in an interval of width $B$. Choosing intervals centered around $x=0$ and $p=0$, respectively, the corresponding probabilities can be determined for any given quantum state $\mid \psi \rangle$ through the integrals of the probability densities,
\begin{equation}
P(\mbox{L}) = \int_{-L/2}^{L/2} |\langle x \mid \psi \rangle|^2 dx
\end{equation}
and 
\begin{equation}
P(\mbox{B}) = \int_{-B/2}^{B/2} |\langle p \mid \psi \rangle|^2 dp.
\end{equation}
Quantum uncertainty imposes a limit on these probabilities whenever the product of the intervals $L$ and $B$ are smaller than $2 \pi \hbar$. As discussed in \cite{Lah86,Bus07}, this uncertainty limit can be expressed as an upper bound of the sum of the two probabilities given by 
\begin{equation}
\label{eq:uncertainty}
P(\mbox{L})+P(\mbox{B}) \leq 1+\sqrt{U},
\end{equation}
where the value of $U$ represents the ratio of the product of $L$ and $B$ to Planck's constant,
\begin{equation}
\label{eq:U}
U = \frac{L B}{2 \pi \hbar}.
\end{equation}
Since $2 \pi \hbar$ describes the phase space area associated with conventional uncertainties, the ratio $U$ defines the suppression of uncertainty achieved by a simultaneous confinement of the particle in L and in B. To achieve a high level of control, low values of $U$ are necessary. At the same time, Eq.(\ref{eq:uncertainty}) shows that values of $U<1$ make it impossible to simultaneously achieve $P(\mbox{L})=1$ and $P(\mbox{B})=1$. However, Eq.(\ref{eq:uncertainty}) also indicates that it is possible to achieve probability sums larger than one. This means that there will be a minimal joint probability of (L and B)=(L,B) since the maximal probability of (L or B) is one, and this probability consists of (L, not B), (B, not L) and (L,B). Adding $P(\mbox{L})$ and $P(\mbox{B})$ counts the probability of (L,B) twice, so the exact joint probability of (L,B) could be obtained by subtracting the value of the probability of (L or B) from this sum. Since the probability of (L or B) is necessarily smaller than one, the minimal joint probability of (L,B) is given by
\begin{equation}
\label{eq:joint}
P(\mbox{L,B}) \geq P(\mbox{L}) + P(\mbox{B}) - 1.
\end{equation}
This equation shows that the highest value of the minimal joint probability $P(\mbox{L,B})$ is obtained from states that saturate the uncertainty relation given in Eq.(\ref{eq:uncertainty})\cite{Lah86,Bus07}.
However, a sufficient joint control of position and momentum can also be obtained with non-optimal states. In the following, the focus will be on the precise mechanisms that result in values close to the uncertainty limit of $\sqrt{U}$. In this context, it is interesting to note that the superposition of rectangular wavefunctions introduced in \cite{Hof17} saturates the uncertainty bound for small values of $U$, indicating the optimized wavefunctions derived in \cite{Lah86,Bus07} converge on such superpositions in the limit of $U \to 0$. It is also interesting to note that the value of $\sqrt{U}$ by which the optimized wavefunctions exceed a probability sum of one is directly related to the overlap between the rectangular wavefunction in position and the rectangular wavefunction in momentum, indicating that the main physical effect that is responsible for the saturation of the uncertainty bound in Eq.{\ref{eq:uncertainty}) is the same interference effect between position and momentum that also proves that quantum particles do not move in straight lines. To clarify the latter effect, the goal of the following discussion will be the analysis of the interference effect between a state component described by an arbitrarily shaped wavefunction localized in position and a state component similarly localized in momentum. Ultimately, the idea behind this analysis is that the quantum interference effect that produces the minimal joint probability $P(\mbox{L,B})$ represents a different causality from the straight line propagation suggested by the operator relation in Eq.(\ref{eq:Heis}), which is demonstrated by the fact that the minimal probabilities that would be associated with $P(\mbox{L,B})$ for motion in a straight line do not show up in the probability distributions of $x(t)$ at later times $t$. 

To describe the control of future positions $\hat{x}(t)$ associated with the probabilities $P(\mbox{L})$ and $P(\mbox{B})$ for initial position and momentum it is necessary to introduce an additional probability $P(\mbox{M})$ that describes the probability of finding the freely propagating particle at a time $t$ in an interval that contains all straight line solutions obtained from combinations of position and momentum from the intervals $L$ and $B$. Specifically, a particle found in the position interval of width $L$ and a momentum interval of width $B$ at time zero should be found in a position interval of width $L+Bt/m$ at any later time $t$. For reasons of symmetry, the most interesting time to consider is $t=mL/B$, where the contributions of the initial intervals are equal. Newton's first law would require that a particle satisfying the initial conditions L and B would also satisfy the condition M, which can be given as
\begin{equation}
|x(t=mL/B)|\leq L.
\end{equation}
We can evaluate the probability $P(\mbox{M})$ that the condition M is satisfied by integrating the probability density at time $t=ml/B$ over the interval of width $2L$ running from $x=-L$ to $x=+L$,
\begin{equation}
P(\mbox{M}) = \int_{-L}^L |\langle x \mid \hat{U}(t=mL/B) \mid \psi \rangle|^2 dx.
\end{equation}
Since the condition M is a necessary consequence of motion in a straight line for particles that satisfy both the spatial condition L and the momentum condition B, it is possible to derive an experimentally testable criterion for Newton's first law based on the probabilities of the conditions L, B, M. This condition is the particle propagation inequality introduced in \cite{Hof17},
\begin{equation}
\label{eq:ppp}
P(\mbox{M}) \geq P(\mbox{L}) + P(\mbox{B}) - 1. 
\end{equation}
As shown in \cite{Hof17}, this inequality can be violated by superpositions of rectangular states of position and momentum, where the quantum interference effect observed at $t=mL/B$ describes a probability that can be about seven percent lower than the minimum given by the right hand side of Eq.(\ref{eq:ppp}). This result appears to be close to the maximal value for the defect probability allowed by the Hilbert space expressions for the probabilities involved. 

For practical reasons, it would be good to know whether the effect observed for rectangular wavefunctions is robust against changes of the waveform. If it is correct that the physical reason for the violation of the inequality in Eq.(\ref{eq:ppp}) is the difference between quantum interference and joint realities, it should be possible to formulate simple quantitative conditions for the violation of the inequality based on only a few characteristic coefficients of the wavefunctions describing the localized components of the state. In the following, the possibility of observing the particle propagation paradox with superpositions of different types of localized wavefunctions will be considered.

\section{Particle propagation paradox for a superposition\\ of Gaussian wavefunctions}
\label{Sec:Gauss}

In order to identify the essential physics responsible for the particle propagation paradox, it is necessary to distinguish the most important quantitative contributions in the integrals that define the probabilities of L, B and M from less relevant details of the specific wavefunctions. Such an identification necessarily corresponds to an approximate description of the wavefunctions. Based on the analysis presented in \cite{Hof17}, the hypothesis examined in the present paper is that the observable violation of Eq.(\ref{eq:ppp}) originates from an interference between a state component localized in position and a state component localized in momentum. If this assumption is valid, it is natural to assume that the momentum distribution of the state component localized in position is much wider than the interval $L$ and the spatial wavefunction within that interval can be approximated by the value at $x=0$. Likewise, the momentum distribution of the state localized in position can be approximated by its $p=0$ value within the interval $B$. These approximations are a natural consequence of the choice of states investigated, and their reliability for values of $U$ well below one should be quite high. However, all of the results obtained with such approximations should be confirmed by a more precise calculation. It may therefore be good to start the discussion with a specific example for which the relevant probabilities can be calculated without any approximations. 

The example of Gaussian wavefunctions presents itself as a very good option for such a precise analysis, since both the integration and the propagation of Gaussian states can be solved without any approximations. We can therefore confirm that the superposition of Gaussian wavefunctions that will be analyzed in more detail in Sec. \ref{Sec:LocalG} does indeed violate Eq.(\ref{eq:ppp}) by calculating the precise probabilities for an appropriate choice of Gaussians. In general, a superposition of Gaussian wavefunctions with different variances $\sigma_1$ and $\sigma_2$ is given by
\begin{eqnarray}
\label{eq:TwoGauss}
\langle x \mid \psi_+ \rangle &=& \frac{1}{\sqrt{2\left(1+\sqrt{\frac{2 \sigma_1 \sigma_2}{\sigma_1^2+\sigma_2^2}}\right)}}\left(\frac{1}{(2 \pi \sigma_1^2)^{1/4}} \exp\left(-\frac{x^2}{4 \sigma_1^2}\right) + 
\frac{1}{(2 \pi \sigma_2^2)^{1/4}} \exp\left(-\frac{x^2}{4 \sigma_2^2}\right)\right).
\end{eqnarray}
For $\sigma_1 \ll \sigma_2$, the Gaussian with variance $\sigma_1$ is localized in position and the Gaussian with $\sigma_2$ is localized in momentum. The momentum distribution will be given by a superposition of Gaussians with variances of $\hbar/(2 \sigma_1)$ and $\hbar/(2 \sigma_2)$. Since the Fourier transform of the two Gaussians is a similar sum of two Gaussians with the same ratio of variances, the state $\mid \psi_+ \rangle$ is symmetric under the Fourier transform that transforms the position representation into the momentum representation and vice versa. Since the variances exchange their roles, the ratio of the momentum scale $B$ and the position scale $L$ is given by $2 \sigma_1 \sigma_2 = \hbar L/B$. The uncertainty suppression factor $U$ must therefore satisfy the relation $4 \pi U \sigma_1 \sigma_2 = L^2$. 

Because of the symmetry of the state under Fourier transforms, the probabilities $P(\mbox{L})$ and $P(\mbox{B})$ are equal. It is a straightforward matter to determine the value of $P(\mbox{L})=P(\mbox{B})$ using the integral of the probability density of the state in Eq.(\ref{eq:TwoGauss}). Here, the goal is to observe a maximal violation of the article propagation inequality in Eq.{\ref{eq:ppp}). As the following analysis will show, a particularly strong violation can be obtained for a combination of $\sigma_1=0.16 L$ and $\sigma_2=22.67 L$. For this set of variances, the right hand side of the particle propagation inequality is given by 
\begin{equation}
P(\mbox{L})+P(\mbox{B})-1 = 0.114569.
\end{equation}
The value of the uncertainty suppression factor for this combination of variances is $U=0.021939$, and the corresponding uncertainty limit for the right hand side of Eq.(\ref{eq:ppp}) given by Eq.(\ref{eq:uncertainty}) would be $\sqrt{U}=0.148118$. The minimal joint probability $P(\mbox{L},\mbox{B})$ for the superposition of Gaussians is therefore not too far from the theoretically possible maximum at this value of the uncertainty suppression $U$. 

The time evolution of Gaussian wavefunctions is well known and merely results in a time dependent change of the variance and the phases that represent correlations between position and momentum. At $t=mL/B$, both quantum state components will have evolved to the same spatial variance of $\sqrt{\sigma_1^2+\sigma_2^2}$. The superposition of the two components will then show up as an interference pattern given by
\begin{eqnarray}
|\langle x \mid \hat{U}(t=m L/B) \mid \psi_+ \rangle|^2 &=& \frac{1}{2\left(1+\sqrt{\frac{2 \sigma_1 \sigma_2}{\sigma_1^2+\sigma_2^2}}\right)} \frac{1}{\sqrt{2 \pi (\sigma_1^2+\sigma_2^2)}} \exp\left(-\frac{x^2}{2 (\sigma_1^2+\sigma_2^2)}\right) \left(1+\cos\left(\phi(x)\right)\right), 
\end{eqnarray}
where $\phi(x)$ represents the quadratic function of $x$ that determines the phase differences between the two components. Although it would be possible to consider the precise interference pattern given by the phase function $\phi(x)$ in an integration of complex Gaussians, it seems to be sufficient to find an upper bound of the probability based on $\cos(\phi)\leq 1$. It is then possible to determine the corresponding upper bound for the probability $P(\mbox{M})$ of finding the particle in the interval M between $x(t)=-L$ and $x(t)=+L$ by integrating the envelope function of the interference pattern. The result is
\begin{equation}
P(\mbox{M}) \leq 0.0628944.
\end{equation}
Clearly, this probability violates the particle propagation inequality given by Eq.(\ref{eq:ppp}). Specifically, the difference between the lower bound of Eq.(\ref{eq:ppp}) and the maximal value of $P(\mbox{M})$ is equal to 0.0516746, a difference of more than five percent. 

The example given in this section demonstrates that a superposition of Gaussians can indeed violate the particle propagation inequality. No approximations were used in the determination of the results, and the precision of the calculations leaves no room for ambiguity. However, the specific combination of variances was not chosen by accident. As will be shown in the following, it is possible to optimize the localization of any form of wavefunction in such a way that a superposition of a state localized in position with a state localized in momentum results in a maximal violation of Eq.(\ref{eq:ppp}). 

\section{Coefficients for localized quantum state components}
\label{Sec:coeff}

In general, it is possible to construct candidates for the violation of Eq.(\ref{eq:ppp}) by using a superposition of a state localized in position with a state localized in momentum. If the localization is sufficiently strong, it is possible to assume that the Fourier transform of the wavefunction will be approximately constant in small intervals around zero. This approximation can be used to determine the probabilities $P(\mbox{L})$, $P(\mbox{B})$, and $P(\mbox{M})$ from only a small number of coefficients that characterize the localization of the wavefunction.

Let us first consider states $\mid \phi_{\mathrm{L}} \rangle$ that are localized in L, so that $P(\mbox{L})$ is close to one and the probability amplitudes $\langle x \mid \phi_{\mathrm{L}} \rangle$ of positions with $|x|\gg L/2$ are negligible. We can then determine the probability of finding the particle in the momentum interval $|p(0)|\ll B/2$ by using the Fourier transform of $\langle x \mid {\mbox{L}} \rangle$. In this Fourier transform, the phase of the contribution at $p(0)=B/2$ for $x=L/2$ is given by $\pi U/2$. For sufficiently small uncertainty suppression factors $U$, the probability amplitudes within the momentum interval $|p|<B/2$ will therefore be approximately constant. The probability of finding the particle in the momentum interval $|p(0)|\ll B/2$ is then 
\begin{equation}
P(B|\phi_{\mbox{L}}) \approx B |\langle p=0 \mid \phi_{\mathrm{L}} \rangle|^2.
\end{equation}   
The probability amplitude of momentum $\langle p=0 \mid \phi_{\mathrm{L}} \rangle$ is determined by an integral of the wavefunction $\langle x \mid \phi_{\mathrm{L}} \rangle$ that can also serve as a measure of localization. It is convenient to define the coherent spread $C$ of the wavefunction as
\begin{equation}
\label{eq:C}
C = \frac{1}{\sqrt{L}} \int_{-\infty}^{+\infty} \langle x \mid \phi_{\mathrm{L}} \rangle dx.
\end{equation}
For a rectangular wavefunction such as the one discussed in \cite{Hof17}, the coherent spread is equal to one, which is the maximal value obtained for any wavefunction with $P(L|\phi_{\mathrm{L}})=1$. The probability of finding the particle in B can now be expressed in terms of the uncertainty suppression factor $U$ and the coherent spread $C$,
\begin{equation}
\label{eq:cross}
P(B|\phi_{\mathrm{L}}) = |C|^2 U.
\end{equation}
The coherent spread $|C|$ can be larger than one for wavefunctions that are not completely localized in the interval $|x|\leq L/2$. To evaluate the lack of localization, we can introduce the statistical mismatch $\eta$, which is defined as the probability of finding the particle outside of the interval $|x|\leq L/2$,
\begin{equation}
\label{eq:eta}
\eta = 1 - \int_{-L/2}^{+L/2} |\langle x \mid \phi_{\mathrm{L}} \rangle|^2 \;dx.
\end{equation}
With this definition, the minimal joint probability $P(\mbox{L},\mbox{B})$ for a state localized in $L$ can be given as
\begin{equation}
\label{eq:local}
P(\mbox{L}|\phi_{\mathrm{L}}) + P(\mbox{B}|\phi_{\mathrm{L}}) - 1 = |C|^2 U - \eta.
\end{equation}
If the state is given by a rectangular wavefunction of width $L$, the minimal probability is equal to $U$. Most other states will achieve lower values because of the trade-off between coherent spread $|C|$ and mismatch $\eta$. 

Eq.(\ref{eq:local}) applies only to states localized in the interval $|x|\leq L/2$. A similar result can be obtained for states localized in the momentum interval $|p|\leq B/2$. As shown in \cite{Hof17}, constructive quantum interferences between two states will enhance both $P(\mbox{L})$ and $P(\mbox{B})$, resulting in a value of the probability sum that exceeds the uncertainty limit suggested by Eq.(\ref{eq:local}). In the following, we will therefore take a look at the effects of an equal quantum superpositions of a state localized in position and a state localized in momentum on the probabilities of finding the particle in the intervals $|x|\leq L/2$ and $|p|\leq B/2$, respectively. 

\section{Quantum interference of position and momentum}
\label{Sec:control}

In a superposition of two non-orthogonal states, the interference term contributes a positive or negative value to the total probability of one for the normalized state, depending on the phase relation between the two states of the superposition. It is therefore convenient to define the global quantum phases of the states in such a way that the inner product of the Hilbert space vectors is a positive real number. This convention will be used throughout the following discussion. 
Since the goal is to increase the probabilities of finding the particle in L and of finding the particle in B, we will now consider constructive interferences of a state $\mid \phi_{\mathrm{L}} \rangle$ localized in $|x|\leq L/2$ and a state $\mid \phi_{\mathrm{B}} \rangle$ localized in $|p|\leq B/2$. The normalized quantum state $\mid \psi_+ \rangle$ is then given by 
\begin{equation}
\mid \psi_+ \rangle = \frac{1}{\sqrt{2 + 2 \langle \phi_{\mathrm{L}} \mid \phi_{\mathrm{B}} \rangle}} \left(\mid  \phi_{\mathrm{L}} \rangle + \mid \phi_{\mathrm{B}} \rangle \right).
\end{equation}  
Since we wish to control both position and momentum equally well, it is natural to choose the same shape of the wavefunction for both states. Specifically, this means that the probability amplitudes are related by
\begin{equation}
\label{eq:sym}
\langle p \mid \phi_{\mathrm{B}} \rangle = \sqrt{\frac{L}{B}} \;\langle x=\frac{L}{B}p \mid \phi_{\mathrm{L}} \rangle.
\end{equation}
The inner product of the two state vectors can be determined by using the approximate Fourier transform of $\langle x \mid \phi_{\mathrm{L}} \rangle$ introduced in the previous section,
\begin{eqnarray}
\label{eq:LB}
\langle \phi_{\mathrm{L}} \mid \phi_{\mathrm{B}} \rangle &=& \int_{-\infty}^{+\infty}
\langle \phi_{\mathrm{L}} \mid p \rangle \langle p \mid \phi_{\mathrm{B}} \rangle \; dp
\nonumber \\ &\approx&  \langle \phi_{\mathrm{L}} \mid p=0 \rangle \int_{-\infty}^{+\infty}\langle p \mid \phi_{\mathrm{B}} \rangle \; dp
\nonumber \\ &\approx& \sqrt{\frac{LB}{2 \pi \hbar}} \left| \frac{1}{\sqrt{L}}\int_{-\infty}^{+\infty}\langle x \mid \phi_{\mathrm{L}} \rangle \; dx \right|^2.
\end{eqnarray}
The localization of $\mid \phi_{\mathrm{L}} \rangle$ in space and the localization of $\mid \phi_{\mathrm{B}} \rangle$ in momentum allow us to approximately determine the inner product as a function of the uncertainty suppression factor $U$ given by Eq.(\ref{eq:U}) and the coherent spread $C$ given by Eq.(\ref{eq:C}),
\begin{equation}
\label{eq:inner}
\langle \phi_{\mathrm{L}} \mid \phi_{\mathrm{B}} \rangle = |C|^2 \sqrt{U}.
\end{equation}
The quantum state overlap is therefore fully determined by the same two coefficients that also determine the value of $P(B|\phi_{\mathrm{L}})$. Comparison with Eq.(\ref{eq:cross}) shows that the overlap $\langle \phi_{\mathrm{L}} \mid \phi_{\mathrm{B}} \rangle$ is exactly $1/\sqrt{U}$ times larger than the probability $P(B|\phi_{\mathrm{L}})$ (or equivalently, $P(\mbox{L}|\phi_{\mathrm{B}})$), indicating that the relative magnitude and importance of interference effects is enhanced by small uncertainty suppression factors $U$.

We can now examine the effect of the interferences on the probabilities of finding the particle in the intervals $|x|\leq L/2$ and $|p|\leq B/2$. For this purpose, it is useful to express the density matrix of the state as a sum of projectors and interference terms,
\begin{equation}
\label{eq:decomp}
\mid \psi \rangle \langle \psi \mid = \frac{1}{2+2\langle \phi_{\mathrm{L}} \mid \phi_{\mathrm{B}} \rangle} \left( \mid \phi_{\mathrm{L}} \rangle\langle \phi_{\mathrm{L}} \mid
+\mid \phi_{\mathrm{B}} \rangle\langle \phi_{\mathrm{B}} \mid
+\mid \phi_{\mathrm{B}} \rangle\langle \phi_{\mathrm{L}} \mid
+\mid \phi_{\mathrm{L}} \rangle\langle \phi_{\mathrm{B}} \mid
\right).
\end{equation}
The contributions of the projectors are already known from the discussion in the previous section. The contribution of the interference terms can be found by integrating the product of the wavefunctions in the corresponding intervals of position or momentum. For the momentum interval, the integral reads
\begin{equation}
\int_{-B/2}^{B/2} \langle p \mid \phi_{\mathrm{B}} \rangle \langle \phi_{\mathrm{L}} \mid p \rangle \; dp 
\approx 
\langle \phi_{\mathrm{L}} \mid p=0 \rangle 
\int_{-B/2}^{B/2} \langle p \mid \phi_{\mathrm{B}} \rangle \; dp. 
\end{equation}
This integral is very similar to the one used to determine the inner product between the state vectors in Eq.(\ref{eq:LB}). However, the integral only runs from $-B/2$ to $B/2$. If the wavefunction $\langle p \mid \phi_{\mathrm{B}} \rangle$ has non-vanishing values outside of the interval $|p| \leq B/2$, it is necessary to evaluate the reduction of the integral. This can be achieved by defining the coherent cross-section $\gamma$ as
\begin{equation}
\label{eq:gamma}
\gamma = \mbox{Re}\left(\frac{\int_{-L/2}^{+L/2} \langle x \mid \phi_{\mathrm{L}} \rangle \; dx}
{\int_{-\infty}^{+\infty} \langle x \mid \phi_{\mathrm{L}} \rangle \; dx}\right). 
\end{equation} 
We can evaluate the statistical contribution of the interference term and express the probability $P(L)$ of finding the particle in the interval $|x|<L/2$ and the probability $P(B)$ of finding the particle in the interval $|p|<B/2$ in terms of the uncertainty suppression factor $U$ and the coefficients $C$, $\eta$ and $\gamma$ that characterize the localization of the states $\mid \phi_{\mathrm{L}} \rangle$ and $\mid \phi_{\mathrm{B}} \rangle$ in the respective position and momentum intervals. The results for the superposition state $\mid \psi_+ \rangle$ read
\begin{equation}
P(\mbox{L}|\psi_+) = P(\mbox{B}|\psi_+) = \frac{1}{2+2|C|^2 \sqrt{U}} \left(1-\eta + |C|^2 U + 2 \gamma |C|^2 \sqrt{U} \right).
\end{equation}
Importantly, the interference term results in an increase of the probabilities that is proportional to the square root of $U$. Since the suppression factor $U$ is much smaller than one, this contribution outweighs the contribution associated with $P(\mbox{B}|\phi_{\mathrm{L}})$ (or equivalently, $P(\mbox{L}|\phi_{\mathrm{B}})$), which is proportional to $U$. 

The minimal joint probability $P(\mbox{L},\mbox{B})$ is given by
\begin{equation}
P(\mbox{L}|\psi_+) + P(\mbox{B}|\psi_+) -1 = \frac{1}{1+|C|^2 \sqrt{U}} \left(|C|^2 U + (2 \gamma -1) |C|^2 \sqrt{U} - \eta \right).
\end{equation}
If $\eta$ is close to zero, it is possible to obtain minimal joint probabilities at small values of $U$, where the main contribution is associated with the interference term. As a result, quantum interference greatly enhances the fraction of particles with both well-defined position and well-defined momentum. Based on this statistical limit, it is then possible to examine how the initial conditions determine the time evolution of position in the extreme quantum mechanical limit. As initially demonstrated in \cite{Hof17}, the results clearly indicate that quantum mechanics modifies the laws of motion by replacing the simple quantitative relation between momentum and the time evolution of position suggested by Eq.(\ref{eq:Heis}) with a qualitatively different notion of causality.

\section{Statistics of particle propagation}
\label{Sec:prop}

The time evolution of the wavefunctions is determined by the unitary operator $\hat{U}(t)$ that represents the general solution of the time dependent Schr\"odinger equation in free space. Since the time evolution of the wavefunction is given by a linear operator, we can look at the time evolution of the two components separately. This is particularly important because the contributions to the probability distributions of position at $t=mL/B$ originate mostly from the very wide momentum distribution for the component $\mid \phi_{\mathrm{L}} \rangle$ localized in position, while the probability distribution for the component $\mid \phi_{\mathrm{B}} \rangle$ localized in momentum is almost unchanged from the initial distribution. As a result, both wavefunctions can be approximated by the waveforms associated with their maximal uncertainties. 

The contribution of the component $\mid \phi_{\mathrm{L}} \rangle$ is initially localized in space and spreads out as a result of its substantial momentum uncertainty. At $t=mL/B$, the spatial wavefunction is almost completely determined by the initial momentum distribution, with a curved wavefront that represents the expected correlations between position and momentum,
\begin{equation}
\langle x \mid \hat{U}(t) \mid \phi_{\mathrm{L}} \rangle =
\exp\left(i \frac{B}{2 \hbar L} x^2 - i \frac{\pi}{4}\right) \sqrt{\frac{B}{L}}
\langle p=\frac{B}{L}x \mid \phi_{\mathrm{L}} \rangle.
\end{equation}
It is possible to simplify this expression by using the position distribution of the state $\mid \phi_{\mbox{B}} \rangle$ instead of the momentum distribution of the state $\mid \phi_{\mathrm{L}} \rangle$, since the two are related by their definitions. Furthermore, the relation between $B$ and $L$ can be expressed in terms of the uncertainty suppression factor $U$, resulting in a more convenient expression of the time evolved wavefunction given by
\begin{equation}
\langle x \mid \hat{U}(t) \mid \phi_{\mathrm{L}} \rangle = 
\exp\left(i \pi \left(\sqrt{U} \frac{x}{L}\right)^2 - i \frac{\pi}{4}\right) 
\langle x \mid \phi_{\mathrm{B}} \rangle.
\end{equation}
At time $t=mL/B$, the initially localized wavefunction has spread out to a shape and width equivalent to the shape and width of the initially delocalized wavefunction described by $\mid \phi_{\mbox{B}} \rangle$. On the other hand, the comparatively precise definition of a momentum of zero by the state $\mid \phi_{\mbox{B}} \rangle$ means that its time evolution at $t=mL/B$ is nearly negligible, as seen my the rather small phase modulation of the momentum components,
\begin{equation}
\langle p \mid \hat{U}(t) \mid \phi_{\mathrm{B}} \rangle = 
\exp\left(i \pi U \left(\frac{p}{B}\right)^2\right) 
\langle p \mid \phi_{\mathrm{B}} \rangle.
\end{equation}
In the momentum representation of the time evolution of $\mid \phi_{\mathrm{B}} \rangle$,
all phase shifts in the interval $|p|\leq B/2$ are smaller than $U$, and hence much smaller than one. It is therefore possible to conclude that the time evolution leaves the quantum state $\mid \phi_{\mathrm{B}} \rangle$ mostly unchanged. Note that this can also be understood as a consequence of energy-time uncertainty: since the low uncertainty of momentum corresponds to a low energy uncertainty, it will take a time considerably larger than $t=mL/B$ until the time evolution significantly changes the Hilbert space vector of the quantum state. Consequently, the spatial wavefunction at time $t$ can be approximated by the initial wavefunction,
\begin{equation}
\langle x \mid \hat{U}(t) \mid \phi_{\mathrm{B}} \rangle \approx 
\langle x \mid \phi_{\mathrm{B}} \rangle.
\end{equation}
For the superposition state $\mid \psi_+ \rangle$, it is essential that the interference between $\mid \phi_{\mathrm{L}} \rangle$ and $\mid \phi_{\mathrm{B}} \rangle$ is described by the local phase difference, which is characterized by a slowly varying phase near the interval M. The probability distribution of $x(t)$ at $t=mL/B$ is given by
\begin{equation}
\label{eq:pattern}
|\langle x \mid \hat{U}(\frac{mL}{B}) \mid \psi_+ \rangle|^2
= \frac{2 |\langle x \mid \phi_{\mathrm{B}} \rangle|^2}{1+\langle \phi_{\mathrm{L}} \mid \phi_{\mathrm{B}}\rangle} \left(\cos\left(\frac{\pi}{2} \left(\sqrt{U} \frac{x}{L}\right)^2 - \frac{\pi}{8} \right)\right)^2.
\end{equation}
Note that the phase shift of $-\pi/8$ is a result of the constructive interference between the two localized state vectors. As a consequence of this phase shift, the interference pattern contributes a non-vanishing probability of $\langle \phi_{\mathrm{L}} \mid \phi_{\mathrm{B}}\rangle$ to the integral over the probability distribution given by Eq.(\ref{eq:pattern}). The width of the interference fringes scales with $L/\sqrt{U}$ while the envelope function of the interference pattern is determined by the position distribution of $\mid \phi_{\mbox{B}} \rangle$.  

The probability $P(\mbox{M})$ of finding the particles in the interval $|x(t)|\leq L$ can be estimated using the upper limit set by constructive interference,
\begin{equation}
P(\mbox{M}) \leq \frac{4}{1+ |C|^2 \sqrt{U}} |C|^2 U.
\end{equation}
This probability is closely related to the result for $P(\mbox{B}\mid \phi_{\mathrm{L}})$ given by Eq.(\ref{eq:cross}) since the probability density at $x(t)=0$ corresponds to the momentum density at $p=0$ for $\mid \phi_{\mathrm{L}} \rangle$, with a maximal enhancement by a factor of two due to constructive interference. An additional factor of two originates from the width of the interval M, which is exactly twice the width of the correspondingly scaled intervals for L and B. Significantly, quantum interference enhances the probability $P(\mbox{M})$ much less than it enhances the probability sum of $P(\mbox{L})$ and $P(\mbox{B})$, resulting in a violation of inequality (\ref{eq:ppp}). The precise amount of the violation can be given in terms of the defect probability,
\begin{equation}
P_{\mathrm{defect}} = P(\mbox{L}) + P(\mbox{B}) -1 - P(\mbox{M}). 
\end{equation}
Using the relations for arbitrary localized states $\mid \phi_{\mbox{L}} \rangle$ and $\mid \phi_{\mbox{B}} \rangle$ derived above, we can find a lower limit for this defect probability in terms of the localization constants that characterize the states. The result reads
\begin{equation}
\label{eq:defect}
P_{\mathrm{defect}} \geq \frac{1}{1+|C|^2 \sqrt{U}}\left((2 \gamma -1) |C|^2 \sqrt{U} - 3 |C|^2 U - \eta  \right).
\end{equation}
This relation generalizes the result obtained for rectangular wavefunctions in \cite{Hof17}. The choice of rectangular wavefunctions in \cite{Hof17} was motivated by the fact that they are completely localized inside their respective intervals, so that the statistical defect is $\eta=0$ and the coherent cross section is $\gamma=1$. Under these conditions, which require that the wavefunction must be zero outside the localization interval, the rectangular wavefunction also maximizes the coherent spread at a value of $C=1$. The defect probability is then given as a function of uncertainty suppression $U$, with
\begin{equation}
\label{eq:Rdefect}
P_{\mathrm{defect}}(\mbox{rectangle}) \geq \frac{\sqrt{U}}{1+ \sqrt{U}}\left(1 - 3 \sqrt{U} \right).
\end{equation}
This result corresponds to the lower limit given in Eq.(13) of \cite{Hof17}, since for $|C|^2=1$, $\sqrt{U}$ is equal to the inner product given by Eq.(\ref{eq:inner}). As discussed in \cite{Hof17}, the maximal value of the lower bound for rectangular wavefunctions is obtained at an uncertainty suppression factor of 0.024, where the defect probability has a minimal value of 0.072. Using the more general relation given by Eq.(\ref{eq:defect}), it is now possible to find the lower bounds for defect probabilities achieved with a wider variety of wavefunctions. 

\section{Localization characteristics of Gaussian wavefunctions}
\label{Sec:LocalG}

Probably the most widely used mathematical description of localized wavefunctions is the Gaussian wavefunction. As discussed in Sec. \ref{Sec:Gauss}, the integration and propagation of Gaussian states can actually be solved without any approximations, since the time evolution of a Gaussian wavefunction in free space retains its Gaussian shape at all times. In the following, we will apply the general formalism for the evaluation of superpositions of states localized in position and in momentum to Gaussians in order to optimize the particle propagation paradox for a superposition of two Gaussians. The localization of the Gaussians in the intervals $|x|\leq L/2$ and $|p|\leq B/2$ can be describes by the coherent spread $C$. Applying the definition of Eq. (\ref{eq:C}) to Gaussian wavefunctions, the wavefunction for a specific value of $C$ can be given as
\begin{equation}
\label{eq:Gauss}
\langle x \mid \phi_{\mathrm{L}} \rangle = \sqrt{\frac{2}{|C|^2 L}} \exp\left(-2 \pi \left(\frac{x}{|C|^2 L}\right)^2 \right).
\end{equation}
The position uncertainty $\sigma_1$ of this Gaussian is equal to $|C|^2/\sqrt{2 \pi}$ times $L/2$, or about $0.2 L$ at $|C|^2=1$. A coherent spread smaller than one therefore guarantees that most of the wavefunction is localized inside the interval of $|x|\leq L/2$. For comparison with Eq. (\ref{eq:TwoGauss}), it may be useful to consider the relation between the position uncertainty of the state localized in momentum and the position uncertainty of the state localized in momentum. The first of these two relations is given by
\begin{equation}
\sigma_1 = \frac{1}{\sqrt{8 \pi}} |C|^2 L.
\end{equation}
As mentioned in Sec. \ref{Sec:Gauss}, the value of the second variance can be derived from the requirement that the ratio between $\sigma_1$ and the momentum uncertainty $\hbar/(2 \sigma_2)$ of the state localized in momentum must be equal to the ratio of the intervals $L/B$. Specifically, the product of the spatial variances is directly related to the uncertainty suppression factor $U$ by
\begin{equation}
\sigma_1 \sigma_2 = \frac{1}{4 \pi U} L^2.
\end{equation}
Therefore, the value of $\sigma_2$ is given by  
\begin{equation}
\sigma_2 = \frac{1}{\sqrt{2 \pi} U |C|^2} L.
\end{equation}
With these relations, it is possibel to derive the corresponding spatial variances $\sigma_1$ and $\sigma_2$ for any combination of uncertainty suppression factor $U$ and squared coherent spread $|C|^2$. As demonstrated in Sec. \ref{Sec:Gauss}, it is then possible to confirm the validity of the general relation between localization coefficients and defect probability given by Eq. (\ref{eq:defect}) using the precise integrations of the corresponding Gaussians. However, the errors tend to be rather small for values of $U$ that are much smaller than one. Since the observation of the particle propagation paradox requires values of $U<1/9$, the difference between the precise calculation and the approximate results used to derive the localization coefficients are negligible throughout the region of interest. It is therefore sufficient to analyze the results in terms of the dependence on the coefficients $U$ and $|C|^2$ given by Eq. (\ref{eq:defect}) without any corrections for the small deviations between the values of $\langle x \mid \psi_+ \rangle$ at $x=0$ and at $x=L/2$, or the values of $\langle p \mid \psi_+ \rangle$ at $p=0$ and at $p=B/2$.

\begin{figure}[th]
\vspace{-1.5cm}
\begin{picture}(500,500)
\put(0,0){\makebox(480,450){
\scalebox{0.8}[0.8]{
\includegraphics{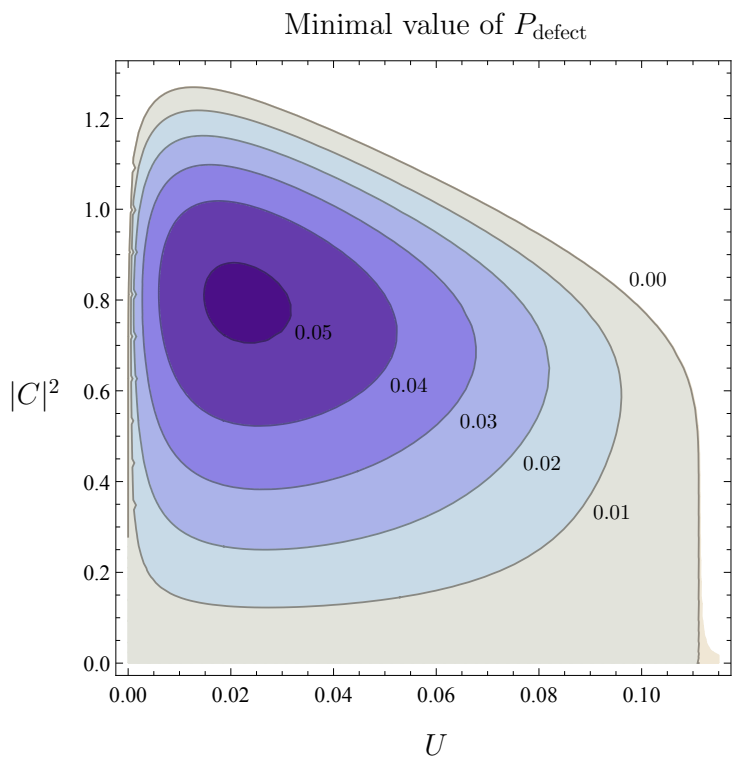}}}}
\end{picture}
\vspace{-5.5cm}
\caption{\label{fig:gauss2D}
Contour plot of the lower bound for the defect probability $P_{\mathrm{defect}}$ as a function of uncertainty suppression factor $U$ and squared coherent spread $|C|^2$ for Gaussian wavefunctions. A violation of the particle propagation inequality by more than five percent is obtained near $U=0.022$ and $|C|^2=0.8$. The outer contour ($P_{\mathrm{defect}}=0.00$) marks the boundary of the inequality violation. 
}
\vspace{0.5cm}
\end{figure}

Since the value of $|C|^2$ completely determines the shape of the Gaussian wavefunction, the remaining coefficients that characterize the localization of the wavefunction can all be expressed as functions of $|C|^2$. 
The statistical mismatch $\eta$ for the Gaussian state is found by solving the integral in Eq.(\ref{eq:eta}),
\begin{equation}
\label{eq:Geta}
\eta = 1 - \mbox{erf} \left(\frac{\sqrt{\pi}}{|C|^2} \right).
\end{equation}
As expected, the statistical mismatch drops to zero for small coherent spreads $|C|$. At a coherent spread of $|C|^2=1$, the value of the statistical mismatch is $\eta(|C|^2=1)=0.01219$, which is still too large to be ignored when compared with the expected values of the defect probabilities. However, $\eta$ rapidly drops to zero as the coherent spread decreases, reaching a value of 0.0017 at $|C|^2=0.8$. 
The coherent cross-section $\gamma$ can be determined by solving the integrals in Eq.(\ref{eq:gamma}),
\begin{equation}
\label{eq:Ggamma}
\gamma = \mbox{erf} \left(\frac{\sqrt{\pi/2}}{|C|^2} \right).
\end{equation}
Note that $\gamma$ is always smaller than $1-\eta$ due to the additional factor if $1/\sqrt{2}$ in the argument of the error function. Specifically, the value at $|C|^2=1$ is $\gamma(|C|^2=1)=0.9237$ and the value at $|C|^2=0.8$ is $\gamma(|C|^2=0.8)=0.9733$. 

\begin{figure}[th]
\vspace{-6.5cm}
\begin{picture}(500,500)
\put(0,0){\makebox(480,450){
\scalebox{0.8}[0.8]{
\includegraphics{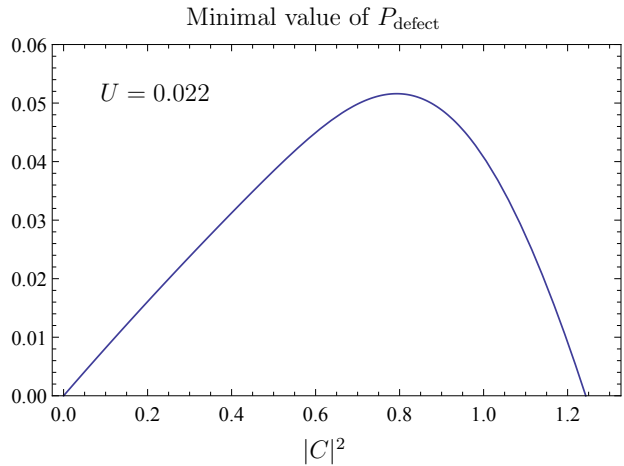}}}}
\end{picture}
\vspace{-4.5cm}
\caption{\label{fig:spread}
Lower bound of the defect probability $P_{\mathrm{defect}}$ as a function of squared coherent spread of the Gaussian wavefunction for the optimal uncertainty suppression factor of $U=0.022$. Initially, the defect probability is proportional to the squared coherent spread $|C|^2$. As the coherent spread increases, an increasing part of the Gaussian wavefunction extends beyond the localization intervals, resulting in a rapid drop of $P_{\mathrm{defect}}$ for $|C|^2>0.8$. 
}
\vspace{0.5cm}
\end{figure}

Eq.(\ref{eq:defect}) indicates that the lower limit of the defect probability increases as $|C|^2$ increases, but decreases with increasing $\eta$ and decreasing $\gamma$. Since an increase of $|C|^2$ represents an increase in the spread of the wavefunction, it naturally results in an increase of the statistical defect $\eta$ and a decrease of the coherent cross section $\gamma$, as shown by Eqs. (\ref{eq:Geta}) and (\ref{eq:Ggamma}). For Gaussians, these effects of the mismatch between the wavefunctions and the intervals start to be relevant in the vicinity of $|C|^2=0.8$ and increase rapidly for higher values of $|C|^2$. Fig. \ref{fig:gauss2D} shows a contour plot of the lower bound achieved for various combinations of uncertainty suppression $U$ and squared coherent spread $|C|^2$. A violation of the particle propagation inequality is obtained for a wide range of combinations between $U=0$ and $U=0.111$, and between $|C|^2=0$ and $|C|^2\approx 1.3$. The maximal violation is obtained at a squared coherent spread of about $|C|^2=0.8$ with an uncertainty suppression factor of about $U=0.022$. The minimal defect probability in this region is slightly below 0.052. As seen in Fig. \ref{fig:gauss2D}, a minimal defect probability of 0.05 can be obtained for squared coherent spreads from $|C|^2=0.7$ to $|C|^2=0.9$ and for uncertainty suppression factors between $U=0.015$ and $U=0.03$. The violation of the particle propagation inequality (\ref{eq:ppp}) is therefore sufficiently robust against small variations of width in both the detection intervals and the wavefunctions. 

Using the relations at the start of the section, we can express the optimal conditions in terms of the spatial Gaussian variances $\sigma_1$ and $\sigma_2$ used in Sec. \ref{Sec:Gauss}. Specifically, $|C|^2=0.8$ corresponds to $\sigma_1=0.1596 L$ and $U=0.022$ results in a value of $\sigma_2=22.667 L$ for the other variance. The values used in the example in Sec. \ref{Sec:Gauss} are therefore close enough to the optimum to permit a direct comparison of the approximate results with the precise integrations of the Gaussian wavefunctions. The calculations show that the minimal defect probability of 0.52 obtained from the localization coefficients of Gaussian states is very close to the precise value of 0.0516746 determined from the exact envelope function of the interference pattern. It is therefore possible to conclude that the approximations used to derive the propagation statistics from the coefficients $|C|^2$, $\eta$ and $\gamma$ are sufficiently reliable to permit optimizations of the particle propagation paradox for a wide variety of different wavefunctions. 

\begin{figure}[th]
\vspace{-6.5cm}
\begin{picture}(500,500)
\put(0,0){\makebox(480,450){
\scalebox{0.8}[0.8]{
\includegraphics{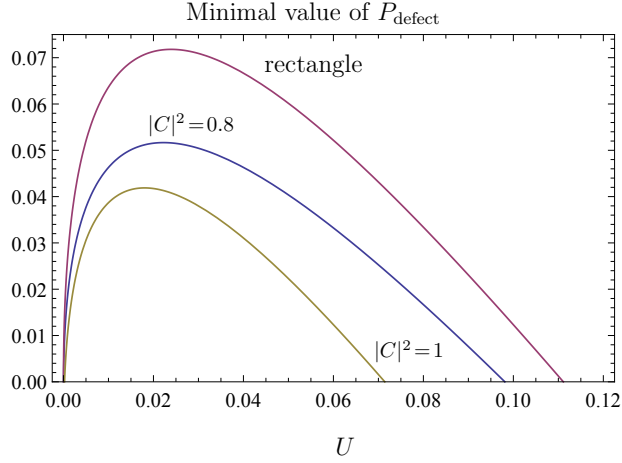}}}}
\end{picture}
\vspace{-4.5cm}
\caption{\label{fig:compare} 
Comparison of defect probability optimization for rectangular wavefunctions (topmost curve), Gaussian wavefunctions with squared coherent spread of $|C|^2=0.8$ (middle curve) and Gaussian wavefunctions with squared coherent spread of $|C|^2=1$ (lowest curve). In all three cases, the defect probability $P_{\mbox{defect}}$ rises sharply at small uncertainty suppression factors, achieving a maximum around $U=0.02$, followed by a gradual drop to zero. 
}
\vspace{0.5cm}
\end{figure}

As indicated above, the optimal value of 0.8 for the squared coherent spread $|C|^2$ originates from the trade-off between the increase of the probability density $P(\mbox{B}|\phi_{\mathrm{L}})=|C|^2 \sqrt{U}$ and the decrease of the localization of the wavefunction in the interval $|x|\leq L/2$ given by the cross-sections $1-\eta$ and $\gamma$. The effect of this trade-off is illustrated for the optimal uncertainty suppression of $U=0.022$ in Fig. {\ref{fig:spread}. The defect probability increases nearly linearly with $|C|^2$ until it starts to level off near the maximal value of 0.052 at $|C|^2=0.8$. Beyond the maximum, there is a rather steep drop in the minimal defect probability, with no violation of the particle propagation inequality beyond $|C|^2\approx 1.25$ where too much of the wavefunction $\langle x \mid \phi_{\mathrm{L}}\rangle$ is located outside of the interval $x \leq L/2$. 

\begin{figure}[th]
\vspace{-6.5cm}
\begin{picture}(500,500)
\put(0,0){\makebox(480,450){
\scalebox{0.8}[0.8]{
\includegraphics{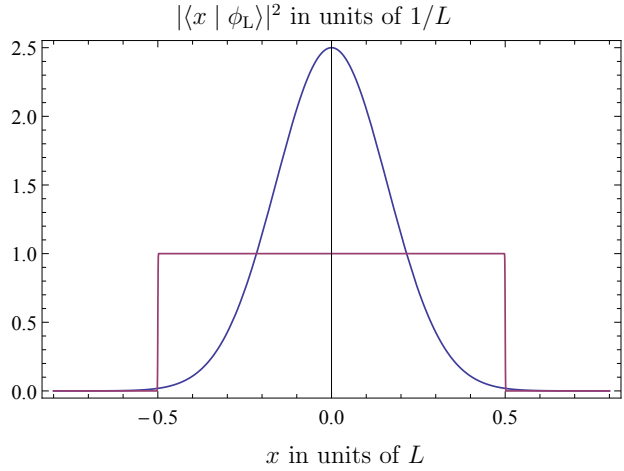}}}}
\end{picture}
\vspace{-4.5cm}
\caption{\label{fig:waveforms} 
Comparison of the probability density of the optimized Gaussian wavefunction with the probability density of the rectangular wavefunction. The optimized Gaussian with a squared coherent spread of $|C|^2=0.8$ is almost completely localized inside the interval of $|x|\leq L/2$, with a statistical mismatch of only $\eta=0.0017$ and a coherent cross-section of $\gamma=0.9733$.
}
\vspace{0.5cm}
\end{figure}

The performance of the Gaussian wavefunctions is compared with the performance of the rectangular wavefunction in Fig. \ref{fig:compare}. The figure shows the dependence of minimal defect probability $P_{\mathrm{defect}}(\mbox{min.})$ on the uncertainty suppression factor $U$ for the rectangular wavefunction discussed in \cite{Hof17}, for the Gaussian with optimal coherent spread of $|C|^2=0.8$, and for the Gaussian with $|C|^2=1$. As expected, the rectangular wavefunction achieves the highest possible defect probabilities for all uncertainty suppression factors. Likewise, the Gaussian with optimal coherent spread $|C|^2=0.8$ outperforms the non-optimized Gaussian with $|C|^2=1$ at all $U$. Comparing the rectangular wavefunction with the non-optimized Gaussian is interesting because they share the same value of coherent spread, $|C|^2=1$. The difference in defect probability is therefore a direct consequence of the statistical defect of $\eta=0.01219$ and the reduced coherent cross-section of $\gamma=0.9237$. The effects of these mismatches between the wavefunctions and the target intervals reduces the observed violation of the particle propagation inequality by about three percent. 

The optimization of defect probabilities results in a well-defined match between the width of a particular wavefunction and the target interval, representing the optimal trade-off between localization in the interval and the overall width of the distribution given by $|\langle x \mid \phi_{\mathrm{L}}\rangle|^2$. The probability distribution of the optimal Gaussian with $|C|^2=0.8$ is compared with the probability distribution of the rectangular state in Fig. \ref{fig:waveforms}. The narrowness of the Gaussian distribution illustrates the importance of fitting most of the wavefunction into the interval between $x=-L/2$ and $x=+L/2$. Based on Eq.(\ref{eq:defect}), it can be concluded that a non-vanishing defect probability can be obtained for any wavefunction with negligible statistical defect $\eta$ and coherent cross-section $\gamma$ close to one, even if the squared coherent spread $|C|^2$ is very low. This means that most wavefunctions can be narrowed down sufficiently to localize them in their target intervals. The essential mechanism that results in the violation of the particle propagation inequality (\ref{eq:ppp}) is the interference between momentum and position that is seen in the interference pattern at time $t=mL/B$.

\section{Quantum limit of causality}
\label{Sec:causality}

It is possible to illustrate the deviation of particle propagation from Newton's first law by comparing the probability distribution at time $t=mL/B$ given by Eq.(\ref{eq:pattern}) with the minimal joint probability $P(\mbox{L},\mbox{B})$ concentrated in the interval $|x(t)|\leq L$. The result for the optimized Gaussian wavefunction is shown in Fig. \ref{fig:paradox}. In this illustration, the magnitude of the violation appears as a ratio of probabilities,
\begin{equation}
\label{eq:ratiodef}
\frac{P(\mbox{M}|\psi_+)}{P(\mbox{L}|\psi_+)+P(\mbox{B}|\psi_+)-1} = 1 - \frac{P_{\mathrm{defect}}}{P(\mbox{L}|\psi_+)+P(\mbox{B}|\psi_+)-1}.
\end{equation}
On both sides of the equation, the denominator is equal to the minimal joint probability $P(\mbox{L},\mbox{B})$ obtained from the initial position and momentum distributions. In the optimized case shown in Fig. \ref{fig:paradox}, this minimal joint probability is equal to 0.115 (0.114569 in the more precise calculation of Sec. \ref{Sec:Gauss}). The upper estimate for the probability $P(\mbox{M})$ is 0.063 (0.0628944 in the more precise calculation of Sec. \ref{Sec:Gauss}). The result is a ratio of less than 0.55, corresponding to a probability density that is nearly half of the density expected from propagation in a straight line. Experimentally, it might therefore be easier to identify the contrast between the expected probability density and the actual probability density at $x(t)=0$. 

If small probability densities can be resolved, the suppression of probability densities near $x(t)=0$ by the ration given in Eq.(\ref{eq:ratiodef}) continues to decrease as the uncertainty suppression factor $U$ drops to zero. Specifically, the minimal joint probability $P(L|\psi_+)+P(B|\psi_+)-1$ drops to zero with $\sqrt{U}$, but the probability $P(M)$ of finding the particle in the interval $|x(t)|\leq L$ at time $t=mL/B$ drops to zero with $U$. As a result, the suppression of probability densities near $x(t)=0$ is actually more extreme at very low values of $U$. If absolute count rates are not an issue, it might thus be worthwhile to explore extremely low values of $U$ to better understand the microscopic causes of this suppression of probabilities. The dependence of the ratio on the uncertainty suppression factor $U$ and the localization coefficients $|C|^2$, $\eta$ and $\gamma$ is given by 
\begin{equation}
\label{eq:ratio1}
\frac{P(M|\psi_+)}{P(L|\psi_+)+P(B|\psi_+)-1} \leq \frac{4 |C|^2 U}{|C|^2 U + (2 \gamma-1)|C|^2 \sqrt{U} - \eta}. 
\end{equation}
As we have seen in the previous section, it is possible to suppress $\eta$ and $1-\gamma$ to zero by decreasing the squared coherent spread $|C|^2$. In this highly localized limit, the ratio of the probabilities in Eq.(\ref{eq:ratio1}) is given by 
\begin{equation}
\label{eq:ratio2}
\frac{P(M|\psi_+)}{P(L|\psi_+)+P(B|\psi_+)-1} \leq \frac{4 \sqrt{U}}{1+\sqrt{U}}. 
\end{equation}
This limit does not depend on $|C|^2$ and is smaller than the limit of one set by the particle propagation inequality (\ref{eq:ppp}) for any uncertainty suppression factor smaller than 1/9. It is therefore possible to violate the particle propagation inequality by any superpositions of wavefunctions that are completely localized inside their respective position and momentum intervals, where the ratio of observed probability $P(M)$ to the minimal probability $P(L,B)$ required by Newton's first law can be suppressed to arbitrarily low values by choosing correspondingly low uncertainty suppression factors $U$. 

\begin{figure}[th]
\vspace{-6.5cm}
\begin{picture}(500,500)
\put(0,0){\makebox(480,450){
\scalebox{0.8}[0.8]{
\includegraphics{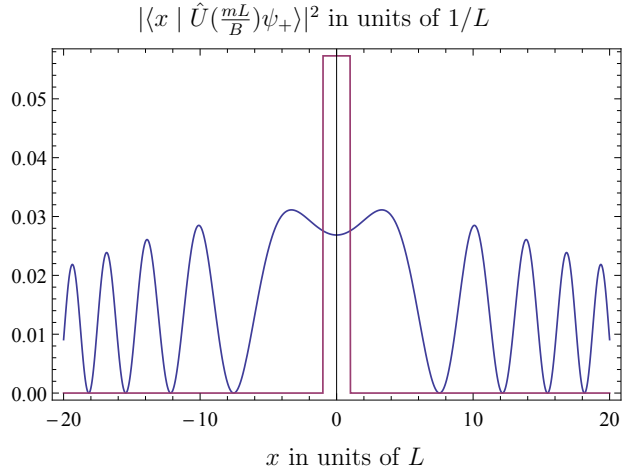}}}}
\end{picture}
\vspace{-4.5cm}
\caption{\label{fig:paradox} Comparison of the probability distribution at time $t=mL/B$ given by Eq.(\ref{eq:pattern}) with the minimal joint probability of $P(L,B)$ concentrated in the interval $|x(t)|\leq L$ for a Gaussian wavefunction with $|C|^2=0.8$ and an uncertainty suppression factor of $U=0.022$. The minimal joint probability is $P(L,B)=0.115$ and the upper bound of the probability $P(M)$ in the interval $|x(t)|\leq L$ is $P(M)<0.063$, for a minimal defect probability of $P_{\mathrm{defect}}=0.052$. The actual value of $P(M)$ is closer to $0.054$, so the actual defect probability is $P_{\mathrm{defect}} \approx 0.061$.
}
\vspace{0.5cm}
\end{figure}

Experimentally, the problem of using highly localized wavefunctions ($|C|^2 \ll 1$) is that the actual probabilities $P(L|\psi_+)$ and $P(B|\psi_+)$ will only be slightly higher than 0.5, making it difficult to obtain sufficiently reliable data about the excess probability $P(L)+P(B)-1$. Specifically, it is necessary to determine the slight increase of probability approximately given by
\begin{equation}
\label{eq:verylocal}
P(L|\psi_+) = P(B|\psi_+) \approx \frac{1}{2}(1 + |C|^2 \sqrt{U}),
\end{equation}
where $|C|^2 \sqrt{U} \ll 1$. 
This difficulty is the reason why the wavefunction localization should be optimized for maximal $P_{\mathrm{defect}}$, and not to the ratio of probability densities near $x(t)=0$. However, the physical origin of the violation of Newton's first law is easier to understand when looking at the dependence of the probability ratio in Eq.(\ref{eq:ratio2}) on uncertainty suppression $U$. If we concentrate on the case of $\eta=0$ and $\gamma=1$, we can scale the interference pattern observed at $t=mL/B$ using the minimal probability density required by Newton's first law, 
\begin{equation}
\label{eq:flatpattern}
\frac{2L |\langle x \mid \hat{U}(\frac{mL}{B}) \mid \psi_+ \rangle|^2}
{P(L|\psi_+)+P(B|\psi_+)-1}
\approx \frac{4 \sqrt{U}}{1+\sqrt{U}}\left(\frac{1}{2}+\frac{1}{2}\cos\left(\pi \left(\sqrt{U} \frac{x}{L}\right)^2 - \frac{\pi}{4} \right)\right).
\end{equation}
In this representation of the probability distribution observed at $t=mL/B$, the upper limit of the ratio of probabilities given in Eq.(\ref{eq:ratio2}) appears as the amplitude of the scaled quantum interference pattern. Importantly, this interference pattern is also responsible for the enhancement of the probabilities $P(L|\psi_+)$ and $P(B|\psi_+)$ beyond 0.5, as explained in Sec. \ref{sec:control}. According to the logic of classical causality, the probability associated with quantum interference should be concentrated exclusively in the interval $|x(t)|\leq L$, resulting in the minimal ratio of one expected for the scaled probability distribution. However, the actual probability densities near $x(t)=0$ are much lower, suggesting that the probability contributed by quantum interference is distributed over a much wider interval. This can indeed be verified by considering the shape of the interference term. The interference is constructive because the integral over the complete interference pattern is positive,
\begin{equation}
\label{eq:width}
\int_{-\infty}^{\infty} \cos\left(\pi \left(\sqrt{U} \frac{x}{L}\right)^2 - \frac{\pi}{4} \right) = \frac{L}{\sqrt{U}}.
\end{equation}
Constructive interference dominates because the phases of the cosine are close to zero in a wide region around $x=0$. Since the values of the cosine in this region are close to one, the integral in Eq.(\ref{eq:width}) defines an effective width of the interference pattern. At $L/\sqrt{U}$, this width can be significantly wider than the width of $2L$ expected for particles with initial position $|x|\leq L/2$ and momentum $|p|\leq B/2$ according to Newtonian laws of motion. It is therefore possible to identify the origin of the violation of Newton's first law in the experimentally observable data: the joint contribution of quantum interference to the position and momentum probabilities $P(L)$ and $P(B)$ is spread out in a quantum interference pattern that is widened by a factor of $\sqrt{U}$ at the time $t=mL/B$ where the joint control of position and momentum would have its maximal effect. We can therefore conclude that quantum interference provides us with a non-trivial modification of the relations between physical properties at different times. The fact that this modification is not dependent on the detailed shape of the wavefunction and is instead given by a very general feature of interferences between low uncertainties in position and low uncertainties in momentum indicates that a better practical understanding of these modifications is possible if we shift the focus of our investigations towards the relation between quantum interference and causality relations between physical properties, as previously suggested in \cite{Hof12,Hof14}.

\section{Conclusions}
\label{Sec:concl} 

Particle propagation is the most accessible example of causality relations in physics, and the idea of controlling the path of a particle by defining its initial position and momentum corresponds to our intuitive understanding of the laws of motion. It is therefore important to understand how these intuitive notions appear in their quantum mechanical limit, where the manipulation and control of physical objects is limited to the preparation of quantum states with uncertain physical properties. In \cite{Hof17}, it was pointed out that the superposition of a quantum state localized in a position interval $L$ and a quantum state localized in a momentum interval $B$ can overcome the conventional uncertainty limit by providing a minimal statistical fraction of particles whose position and momentum is defined by both $L$ and $B$, so that the uncertainty of the fraction of the ensemble of particles described by the superposition can be below the uncertainty limit of $2\pi\hbar$ by an arbitrary uncertainty suppression factor of $U$. If it was correct to assume that the combination of position and momentum determined the future positions of the particles according to Newton's laws of motion or any similar geometric relation, we would have to find this minimal fraction of particles within a corresponding spatial interval at any later time $t$. Interestingly, quantum mechanics instead describes the effects of superposition as an interference pattern, where the additional probability of constructive quantum interferences between $\mid \phi_{\mathrm{L}} \rangle$ and $\mid \phi_{\mathrm{B}} \rangle$ appears as an oscillation with a quadratic phase. As shown by Eq.(\ref{eq:width}), this interference pattern spreads out the additional probability over an interval that is about $1/(2 \sqrt{U})$ times wider than the interval determined by applying Newton's laws of motion to the position and momentum intervals. 

The effect that quantum interference spreads out the pattern of probability associated with the simultaneous definition of position and momentum is independent of the specific wavefunction used to describe the respective definitions of position and momentum. However, care must be taken to quantify the fractions of the wavefunctions that lie outside of the localization interval. As shown above, this can be done by evaluating the statistical mismatch $\eta$ as defined by Eq.(\ref{eq:eta}) and the coherent cross section $\gamma$ as defined by Eq.(\ref{eq:gamma}). In addition, the effective width of the wavefunction can be evaluated using the coherent spread $C$ as defined by Eq.(\ref{eq:C}). It is then possible to show that the particle propagation inequality (\ref{eq:ppp}) which defines the statistical limit for propagation in a straight line can be violated by superpositions of any kind of wavefunctions localized in position and momentum, respectively, with the quantitative value of the defect probability $P_{\mathrm{defect}}$ given as a function of the wavefunction localization characteristics in Eq.(\ref{eq:defect}). As noted above, it is usually possible to arbitrarily reduce the statistical mismatch $\eta$ and the coherent mismatch $1-\gamma$ at the expense of lowering the quadratic coherent spread $|C|^2$, so that violations of the particle propagation inequality (\ref{eq:ppp}) can always be achieved in the limit of very low uncertainty suppression factors $\hat{U}$. The detailed optimization of Gaussian wavefunctions shows that defect probabilities of more than five percent can be obtained under conditions that are very similar to the ones for rectangular wavefunctions given in \cite{Hof17}. In particular, the optimal uncertainty suppression factor of $U=0.022$ is rather close to the optimal factor of $U=0.024$ obtained for rectangular wavefunctions. 

The analysis above provides the necessary theoretical tools to analyze the localization of wavefunctions with regard to specific intervals of position or momentum. This approach may also be useful in other contexts where continuous variables need to be adapted to statistical arguments about well-defined measurement outcomes. In general, quantum interferences describe necessary corrections to classical laws of causality, and this modification is expressed mathematically in the interference terms between the non-orthogonal states that represent complementary conditions of control such as position and momentum. It may be interesting to note that the mathematical expression of this interference term also appears in weak measurements, and hence in the analysis of quantum paradoxes using weak values \cite{box,Har92,Res04,Lun09,Yok09,Aha13,Den14,Hof15}. The violation of the particle propagation inequality discussed above thus establishes a connection between the rather abstract logic of of quantum paradoxes in few level systems and the more intuitive concept of motion in continuous spaces. This connection would seem to support the idea that quantum mechanics is essentially about a modification of the deterministic relations between physical properties, as suggested in \cite{Nii18}. Specifically, the discussion above shows how quantum interferences modify the physical meaning of the operator relation given in Eq.(\ref{eq:Heis}). Operators and their expectation values tend to obscure the essential role of quantum coherence and quantum phase in the definition of causality and the related means of control. For a complete understanding of quantum physics, it may be necessary to investigate the physics described by quantum phases in more detail, especially with regard to the definition of dynamics and its classical limit \cite{Hof16,Hib18}. 
The detailed analysis of the interference pattern at time $t=mL/B$ in Sec. \ref{Sec:causality} might be seen as a first step towards such an investigation, since it 
not only identifies the origin of the violation of the particle propagation inequality (\ref{eq:ppp}), but also points the way to a constructive explanation of causality and control in the quantum limit. Significantly, the maximal level of microscopic control is always achieved by quantum interference, which provides the means for statistical control beyond any conventional uncertainty limits. However, this higher level of control reveals fundamental differences in the actual causality relations described by quantum phases and the classical limit described by differential geometry. It is essential to identify these differences and to develop a better understanding of the relation between the correct quantum mechanical description of motion and its approximate representation in classical physics. The quantitative demonstration of deviations from Newton's first law demonstrates that it is indeed possible to investigate these differences based on experimentally testable criteria, and could therefore point the way towards a better understanding of the fundamental laws of physics.

\section*{Acknowledgment}
This work has been supported by CREST, Japan Science and Technology Agency.

\end{document}